\documentclass[aps,prb,twocolumn,superscriptaddress,showpacs,preprintnumbers,amsmath,amssymb,floatfix]{revtex4}
\usepackage{graphicx}
\usepackage{dcolumn}
\usepackage{bm}
\graphicspath{{./figs/}}

\begin{document}

\title{Wavepacket scattering on graphene edges in the presence of a (pseudo) magnetic field}

\author{D. R. da Costa}
\author{A. Chaves}
\author{G. A. Farias}
\affiliation{Universidade Federal do Cear\'a, Departamento de
F\'{\i}sica Caixa Postal 6030, 60455-760 Fortaleza, Cear\'a, Brazil}
\author{L. Covaci}
\affiliation{Department of Physics, University of Antwerp, Groenenborgerlaan 171, B-2020 Antwerp, Belgium}
\author{F. M. Peeters}
\affiliation{Department of Physics, University of Antwerp, Groenenborgerlaan 171, B-2020 Antwerp, Belgium}
\affiliation{Universidade Federal do Cear\'a, Departamento de
F\'{\i}sica Caixa Postal 6030, 60455-760 Fortaleza, Cear\'a, Brazil}

\begin{abstract}
The scattering of a Gaussian wavepacket in armchair and zigzag graphene edges is theoretically investigated by numerically solving the time dependent Schr\"odinger equation for the tight-binding model Hamiltonian. Our theory allows to investigate scattering in reciprocal space, and depending on the type of graphene edge we observe scattering within the same valley, or between different valleys. In the presence of an external magnetic field, the well know skipping orbits are observed. However, our results demonstrate that in the case of a pseudo-magnetic field, induced by non-uniform strain, the scattering by an armchair edge results in a non-propagating edge state.
\end{abstract}

\maketitle

\section{Introduction}

Due to its unique electronic properties, graphene has become a topic of intensive study in recent years. Within the low energy approximation for the tight-binding Hamiltonian of graphene, electrons behave as massless Dirac fermions, with a linear energy dispersion. \cite{CastroNetoReview} This leads to a plethora of interesting physical phenomena, ranging from Klein tunneling and other quasi-relativistic effects \cite{Katsnelson, Maksimova} to the existence of new types of electron degrees of freedom, namely, the pseudo-spin, related to the distribution of the wave function over the carbon atoms belonging to the different triangular sub-lattices composing the graphene hexagonal lattice, and the presence of two inequivalent electronic valleys, usually labeled as K and K', in the vicinity of the gapless points of the energy spectrum of graphene.  

Recent papers studied the scattering of electrons by edges \cite{Yang} and defects \cite{Gunlycke} in graphene, both theoretically \cite{Massoud} and experimentally \cite{Chen}. Armchair and zigzag are the two types of edges which are most frequently considered in the study of graphene ribbons, although other types of terminations exist due to edge reconstruction, which has been demonstrated both theoretically \cite{TheoryReconstruction} and experimentally \cite{ExpReconstruction1,ExpReconstruction2,ExpReconstruction3}. Even so, the edge reconstruction effect strongly depends on how the nanoribbon is made: normally, it occurs when the technique used to fabricate the nanoribbon is based on a mechanism that drives the system to thermodynamic equilibrium. According to the continuum (Dirac) model, armchair edges in finite graphene samples lead to a boundary condition that mixes the wavefunctions of K and K' valleys, whereas a zigzag edge appears in the Dirac theory of graphene as a separate boundary condition for the wavefunctions of each valley. \cite{Akhmerov, Mohammad} This suggests that electrons reflected by a graphene edge would exhibit inter-valley scattering only in the armchair case, whereas reflection by a zigzag edge would produce scattering inside the same Dirac valley. This prediction was confirmed by recent experiments, \cite{Park} where inter-valley scattering by armchair edges was even shown to be very robust in the presence of defects. The inter- and intra-valley scattering possibilities are schematically illustrated in Fig. 1(a), which shows K and K' Dirac cones in the reciprocal space of graphene.

Besides its singular electronic properties, graphene also exhibits interesting mechanical properties, as it can support strong elastic stretch. This provides us with the new possibility to tune the electron properties in graphene through strain engineering. \cite{CastroNetoStrain,new1, new2,new3,new4,new5,new6,new7,new8,new9,new10} In fact, it has been demonstrated recently that electrons in a strained graphene lattice behave as if they were under an external magnetic field, which points towards opposite directions in the K and K' valleys, so that the time reversal symmetry of the system as a whole is preserved. \cite{Guinea} Such fields were experimentally observed recently, when measurements of the energy states in a graphene bubble revealed a Landau level-like structure corresponding to an external magnetic field of $\approx$ 300 T. \cite{CastroNeto300T} By designing non-uniform strain fields in a graphene sheet, one is able to produce a uniform pseudo-magnetic field for electrons. \cite{GuineaStrain}
\begin{figure}[!h]
\centerline{\includegraphics[width=\linewidth]{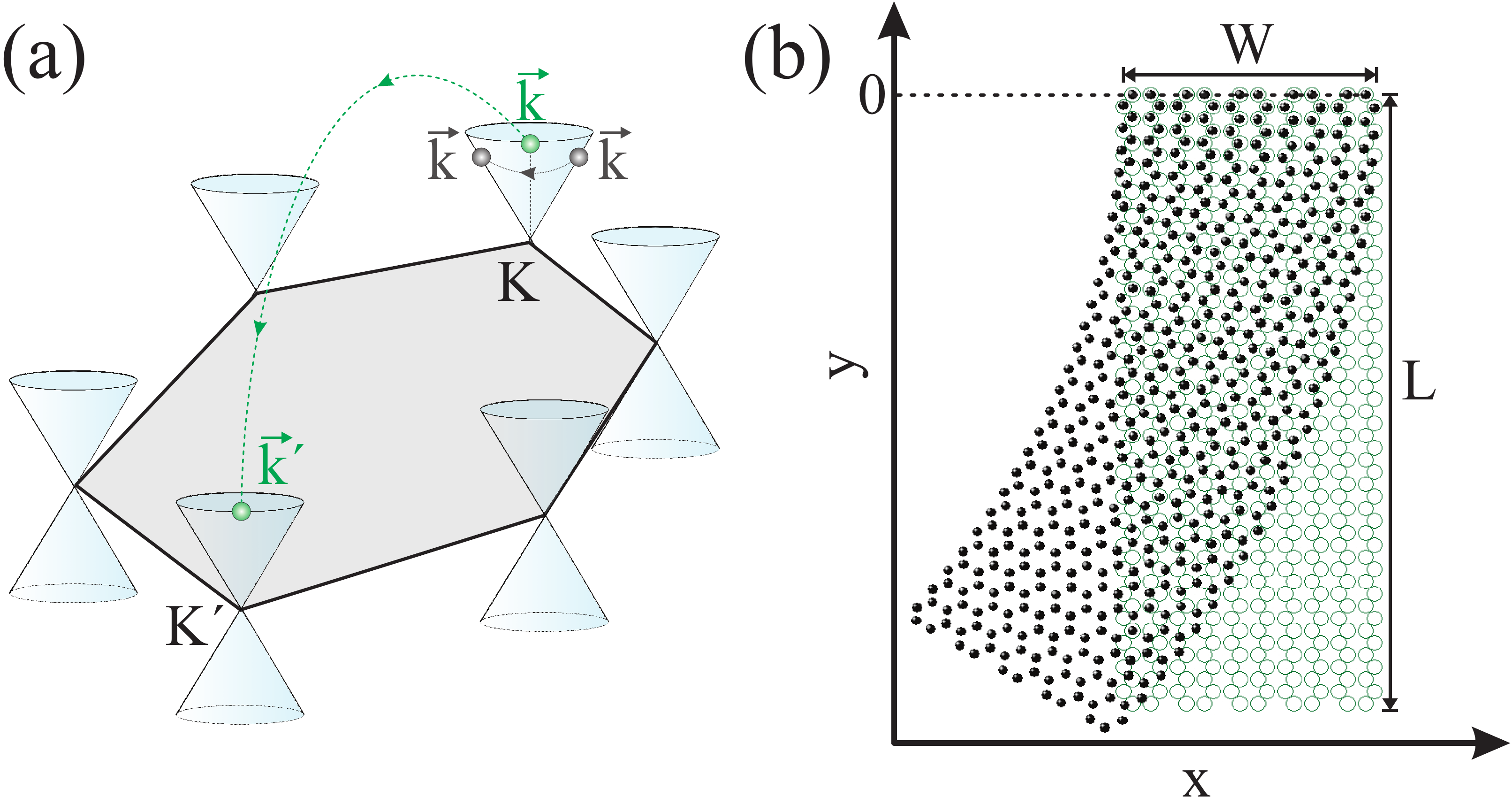}}
\caption{(Color online)(a) Dirac cones of graphene, along with an illustrative scheme of the inter- (green circles) and intra-(gray circles) valley scattering. (b) Sketch of the strained graphene sample considered in this work, where the (open green) full black circles represent the (un)strained case. The upper boundary is set as the $y$ = 0 axis for convenience.} \label{fig:sketch}
\end{figure}

The aim of this paper is two fold - we use wavepacket dynamics calculations: (i) to investigate electron reflection by armchair and zigzag edges in a finite graphene sample, assumed to be made by cutting a graphene monolayer, such that no edge reconstruction is expected to occur at room temperature, where our results demonstrate the possibilities of inter- and intra-valley scattering, depending on the type of edge, and (ii) to study the influence of an external magnetic field and a non-uniform strain distribution on the electron trajectories in these systems. We compare the features observed for electrons under a perpendicular external magnetic field with those seen with a pseudo-magnetic field. Fig. 1(b) shows a sketch of the graphene flake considered in our calculations, where the open (green) circles illustrate the unstrained sample and the closed (black) circles illustrate the strained one. Such a non-uniform strain field was suggested by Guinea \textit{et al.} \cite{GuineaStrain} and was shown to exhibit an almost uniform pseudo-magnetic field.

All the calculations were done within the tight-binding description of graphene, using the time-evolution method developed in Ref. \onlinecite{AndreyStrain}. As we are not restricting ourselves to a single Dirac cone in our model, the scattering between Dirac cones by armchair edges will appear naturally. Notice that Fig. 1(b) is just an illustrative scheme of our system, where the number of atoms was reduced in order to help its visualization. Besides, the sample shown in Fig. 1(b) is a ribbon, which improves the visualization of the strained case. However, the actual flake considered in our calculations has 1801$\times$2000 atoms, which looks more like a rectangle, rather than a ribbon and corresponds to a flake with dimensions of about 426$\times$221 $nm^2$. Such a large flake is necessary to isolate each reflection of the wavepacket on a single edge, as we need to consider a large packet in order to avoid dispersion. \cite{AndreyStrain, Maksimova, Rusin} 

\begin{figure}[!b]
\centerline{\includegraphics[width=0.8\linewidth]{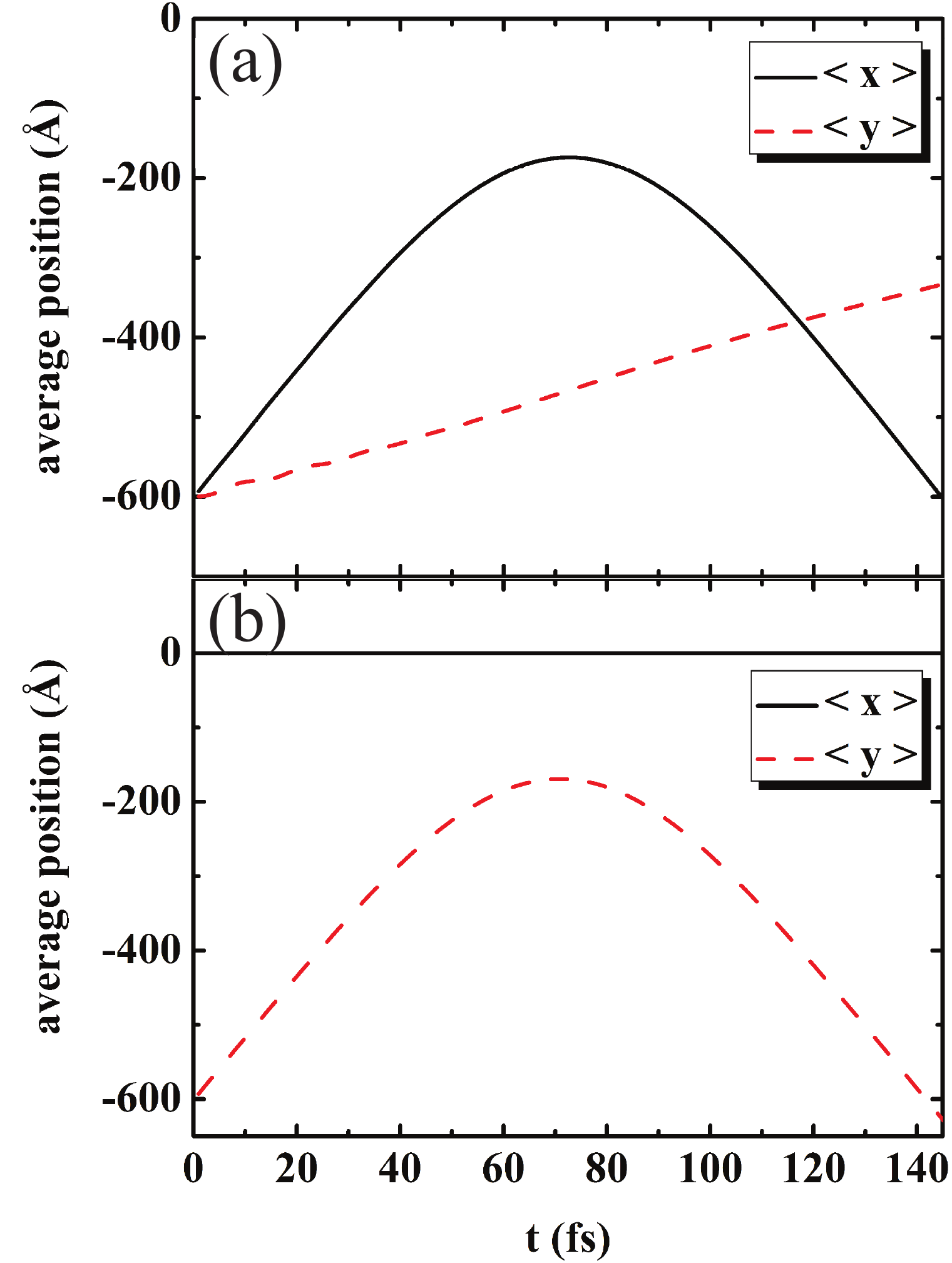}}
\caption{(Color online) Average position of a gaussian wavepacket of width $d = 300$ \AA\, in an unstrained graphene flake, in the absence of external magnetic fields, as a function of time. (a) Horizontal propagation of a wavepacket with $\vec k = (0.03$\AA\,$^{-1}, 4\pi\big/3\sqrt{3}a)$, and its consequent reflection by a zigzag edge. In this case the origin of the system is shifted, so that the right zigzag edge is set as the $x = 0$ axis. The wavepacket starts at $(x_0,y_0) = (-600$\AA\,$, -600 $\AA\,$)$, and exhibits also a slow drag towards the upper edge, due to Zitterbewegung effects. (b) Vertical propagation of a wavepacket with $\vec k = (0, 4\pi\big/3\sqrt{3}a + 0.02$\AA\,$^{-1})$, starting at $(0, -600$\AA\,$)$, exhibiting reflection by the upper armchair border. In this case, the sample is not shifted, \textit{i.e.} the upper edge is at the $y = 0$ axis, as sketched in Fig. 1(b).} \label{fig:centmassunstrained}
\end{figure}

\section{Edge-dependent scattering}

Let us first analyze the wavepacket reflection by zigzag and armchair edges in a plain graphene sample, \textit{i.e.} in the absence of magnetic fields and strain. The initial wavepacket $\Psi(x,y)$ is taken as a circularly symmetric gaussian distribution of width $d$, multiplied by a plane wave with wave vector $\vec k = (k_x, k_y)$ and a pseudo-spinor $\sigma = \left(A , B \right)^T$
\begin{eqnarray}
\Psi(\vec{r}) &=& \frac{1}{d\sqrt{2\pi}}
\left(\begin{array}{c}
A\\
B
\end{array}\right)\nonumber\\
& &\times\exp\left[-\frac{(x-x_0)^2 + (y-y_0)^2}{2d^2} + i\vec{k}.\vec{r}\right]
.\end{eqnarray}

The pseudo-spinor in our model is simulated by defining a multiplication factor in the wave function, which assumes different values for sites belonging to the $A$ and $B$ sub-lattices. Notice that for low energy electrons in graphene, the tight-binding Hamiltonian can be approximated as $H = v_F\hbar \vec k \cdot \vec \sigma$, where $v_F$ is the Fermi velocity, so that the propagation velocity vector in the Heisenberg picture is given by $d\vec x/dt = -[\vec x, H]i/\hbar = v_F\vec{\sigma}$. Hence, the pseudo-spin polarization of the wavepacket plays an important role in defining the direction of propagation. As the upper and right edges of the flake are of armchair and zigzag type, respectively, we consider $\sigma = \left(1 , i \right)^T$, \textit{i.e.} propagation in the $y$-direction, in order to observe wavepacket scattering on the upper armchair edge, and $\sigma = \left(1 , 1 \right)^T$, \textit{i.e.} propagation in the $x$-direction, for scattering on the right zigzag edge. The initial wave vector $\vec k$ is taken in the vicinity of the Dirac point $\vec{K} = (0, 4\pi\big/3\sqrt{3}a)$, where $a = 1.42$\AA\, is the inter-atomic distance. At each time step, we calculate the average values $\langle x \rangle = \int_{-\infty}^{\infty}x|\Psi|^2dxdy$ and $\langle y \rangle = \int_{-\infty}^{\infty}y|\Psi|^2dxdy$, in order to track the wavepacket trajectory in real space. Besides, a fast Fourier transform (FFT) of the wavepacket is taken at each time step, in order to track its scattering in reciprocal space. 

The average positions $\langle x \rangle$ (black solid) and $\langle y \rangle$ (red dashed) are shown in Fig. 2 as a function of time, for a wavepacket propagating in the $x$ ($y$) direction, towards the right zigzag (upper armchair) edge of the sample, and being reflected by this edge back to its initial position. Figs. 2(a) and (b) correspond to zigzag and armchair reflections, respectively. The wavepacket starts at 600 \AA\, from the sample edge and reaches the edge at $t \approx 70$ fs in both cases. Due to the finite width of the packet ($d$ = 300 \AA\,), its center of mass never reaches the border \cite{AndreyStrain}, so that $\langle x \rangle$ or $\langle y \rangle$ start to exhibit backscattering when they are still $\approx 150$ \AA\, far from the edge. Notice that the motion in the $y-$direction shown in Fig. 2(b) is perfectly vertical, \textit{i.e.} $\langle x \rangle = 0$ during the whole propagation. However, this is not the case for propagation in the $x-$direction as apparent in Fig. 2(a), which is not perfectly horizontal, \textit{i.e.} $\langle y \rangle$ does not stay the same, as the wavepacket slowly drags towards larger $y$ during propagation. This effect is a manifestation of the zitterbewegung, as discussed in detail in Ref. \onlinecite{AndreyStrain}. Although we did not manage to construct a wavepacket that propagates perfectly horizontal, avoiding such a vertical drag, this effect does not interfere in our results and conclusions, as our analysis of scattering on the zigzag edge depends only on the horizontal component of motion.

Once we know the instant when the wavepacket is reflected by the graphene edge in real space, at that moment we analyze what happens in reciprocal space. Fig. 3(a) shows the lines (red dashed) in reciprocal space along which we will take the wave functions. The contour plots in Fig. 3(b) illustrate the wave function along the horizontal line $(i)-(ii)$ depicted in Fig. 3(a) in reciprocal space, as time elapses, in the case of $x$-direction propagation and, consequently, zigzag edge reflection. For such a propagation direction, we assumed the initial wave vector as $\vec k = (0.03$\AA\,$^{-1}, 4\pi\big/3\sqrt{3}a)$. Therefore, the initial wavepacket (at $t = 0$) has a peak around $k_x^{i-ii} = 0.03$ \AA\,$^{-1}$. This peak is conserved until the wavepacket starts to be reflected by the right zigzag edge, when interference patterns start to show up. At $\approx 70$ fs, a peak at $k_x^{i-ii} = -0.03$ \AA\,$^{-1}$ starts to appear, while the former peak at $k_x^{i-ii} = 0.03$ \AA\,$^{-1}$ smoothly decays. This is indeed the instant when the wavepacket is reflected by the zigzag edge in real space, as shown in Fig. 2(a). As time elapses, the wavepacket ends up only around $k_x^{i-ii} = -0.03$ \AA\,$^{-1}$. This is direct evidence of intra-valley scattering as schematically illustrated in Fig. 1(a).
\begin{figure}[!h]
\centerline{\includegraphics[width=1.0\linewidth]{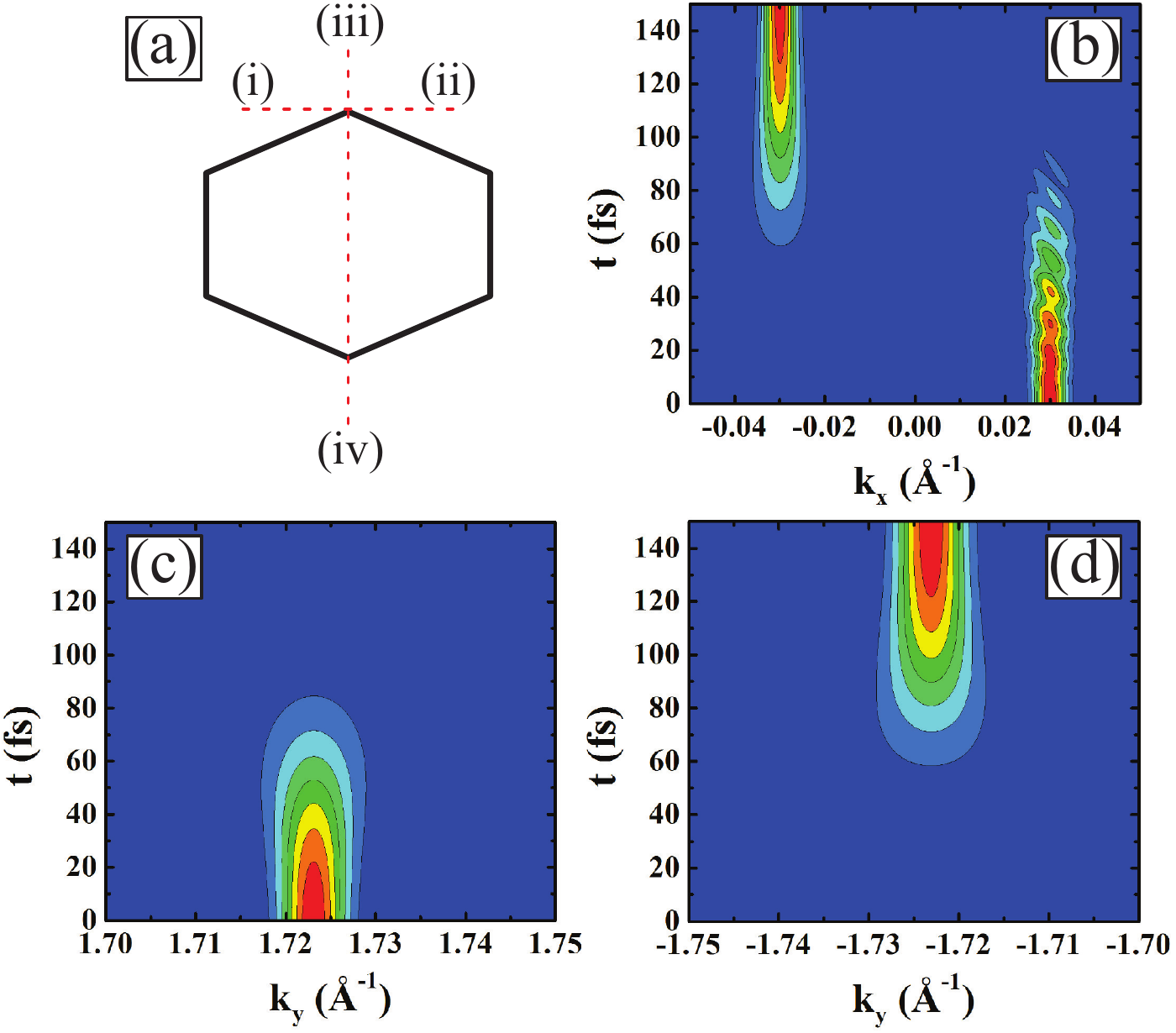}}
\caption{(Color online) Time evolution of the wavepacket in reciprocal space corresponding to the situations shown in Fig. 2. (a) Illustrative scheme of the lines in reciprocal space along which Fourier transform of the wave functions are taken. For the propagation in the horizontal direction (see Fig. 2(a)), we consider $\vec k = (0.03$\AA\,$^{-1}, 4\pi\big/3\sqrt{3}a)$. The time evolution of the wave function along the $(i)-(ii)$ line of reciprocal space is shown in (b) as contour plots. For vertical propagation (see Fig. 2(b)), we consider $\vec k = (0, 4\pi\big/3\sqrt{3}a + 0.02$\AA\,$^{-1})$. The time evolution of the wave function along the $(iii)-(iv)$ line of reciprocal space is shown as contour plots in (c) and (d), corresponding to different ranges of $k_y^{iii-iv}$.} \label{fig:scatteringinkspace}
\end{figure} 

Figs. 3(c) and (d) show the wave function in reciprocal space taken along the vertical line $(iii)-(iv)$ depicted in Fig. 3(a) as time elapses, for vertical propagation and, consequently, armchair edge reflection. For propagation in the $y$-direction, we consider $\vec k = (0, 4\pi\big/3\sqrt{3}a + 0.02$\AA\,$^{-1})$, so that the wavepacket initially exhibits a peak around $\approx 1.723$ \AA\,$^{-1}$, as shown in Fig. 3(c). This peak is preserved up to $t \approx 70$ fs, when the wavepacket is scattered by the upper armchair edge (see Fig. 2 (b)) and the amplitude of the peak starts to decrease. Meanwhile, another peak appears around $k_y^{iii-iv}\approx -1.723$ \AA\,$^{-1}$, which is located in the K' valley, as shown in Fig. 1(a). The inter-valley scattering situation is illustrated by the green circles in Fig. 1(a) which is clearly observed in reciprocal space.
\begin{figure}[!h]
\centerline{\includegraphics[width=1.0\linewidth]{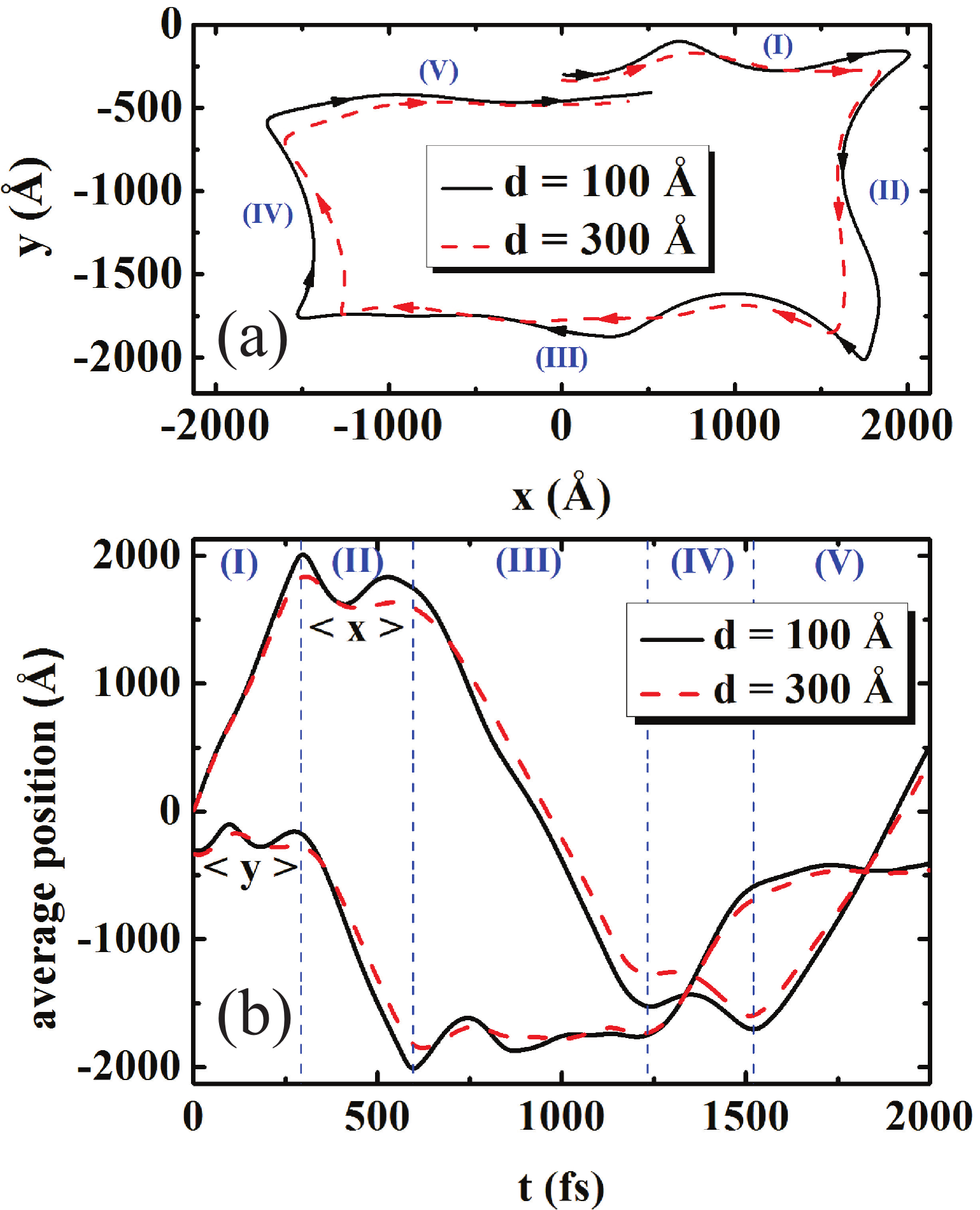}}
\caption{(Color online) Trajectories drawn by $\langle x \rangle$ and $\langle y \rangle$ for a gaussian wavepacket in the presence of an external magnetic field $\approx 5$ T, propagating close to the edges in a rectangular graphene flake, within a $t = 2000$ fs propagation time. The arrows indicate the direction of propagation. Different values of the initial wavepacket width $d$ are considered. The edges of this panel are placed at the positions of the actual edges of the sample. (b) Average values of the wavepacket position $\langle x \rangle$ and $\langle y \rangle$ as a function of time for the trajectories drawn in (a). Different parts of the trajectory in (a) were labeled from (I) to (V), and the time intervals where they occur are delimited by the vertical lines in (b).} \label{fig:trajexternal}
\end{figure}

\section{Skipping orbits}

Let us now investigate the trajectory of a wavepacket in the presence of an external magnetic field, while it undergoes reflection at the edges of our rectangular graphene flake. We consider the same conditions as in Fig. 2(a), \textit{i.e.} the wavepacket in this case moves to the right, being thus pushed to the upper armchair edge by the Lorentz force due to the perpendicular magnetic field. The trajectory drawn by $\vec r = (\langle x \rangle, \langle y \rangle)$ for such a packet in the $xy-plane$ after a $t = 2000$ fs propagation, under a $\approx$5 T field, is shown in Fig. 4(a), where skipping orbits are clearly observed, \cite{skipping} coming from the successive reflections at the borders of the system, followed by ciclotronic semi-circles, as one would expect from such a scattering problem. The arrows indicate the direction of propagation, and the edges of the figure are set to be exactly at the position of the edges of the graphene flake. In order to help their analysis, the trajectories where divided into four regions, labeled from (I) to (V). Fig. 4(b) shows $\langle x \rangle$ and $\langle y \rangle$ separately as a function of time, where one verifies \textit{e.g.} the attachment of the packet to the upper edge ($\langle y \rangle$ close to $y = 0$ in region (I)), followed by a decrease in $\langle y \rangle$, when it attaches to the right edge ($\langle x \rangle$ close to $x = 2100$ \AA\,, in region (II)), and its further attachment to the bottom edge of the sample ($\langle y \rangle$ close to $y = - 2100$ \AA\, in region (III)). As previously mentioned, due to the finiteness of the packet width, the trajectory as described by $(\langle x \rangle, \langle y \rangle)$ does not reach the edges of the system. Besides, the wavepacket disperses as time elapses, which distorts the trajectory as compared to the one obtained by classical ballistic motion. \cite{AndreyStrain} Even so, the main conclusion one draws from this result is quite clear: as well as in ordinary systems with confined Schr\"odinger particles, \cite{skipping2} electrons in graphene under external magnetic fields exhibit a skipping orbit pattern when propagating close to the edges of the sample. We performed calculations for different wavepacket widths $d = 100$ \AA\, and 300 \AA\,, and the results lead to the same qualitative conclusion, differing only by the distance the wavepacket may reach the edge. There is, however, an important difference between these skipping orbits and those in ordinary Schr\"odinger systems, namely, the wavepacket in this case may scatter not only between momentum states with opposite signs within the same valley, as usual, but they can also scatter from one valley to another, depending on the type of edge, as we demonstrated in Fig. 3. Nevertheless, the effect of an external magnetic field on electron states belonging to both valleys is the same, therefore, there is no detectable manifestation of inter-valley scattering in this situation. This is not the case when, instead of an external magnetic field, we consider a strain induced pseudo-magnetic field, as we will demonstrate in what follows. 

\begin{figure}[!h]
\centerline{\includegraphics[width=\linewidth]{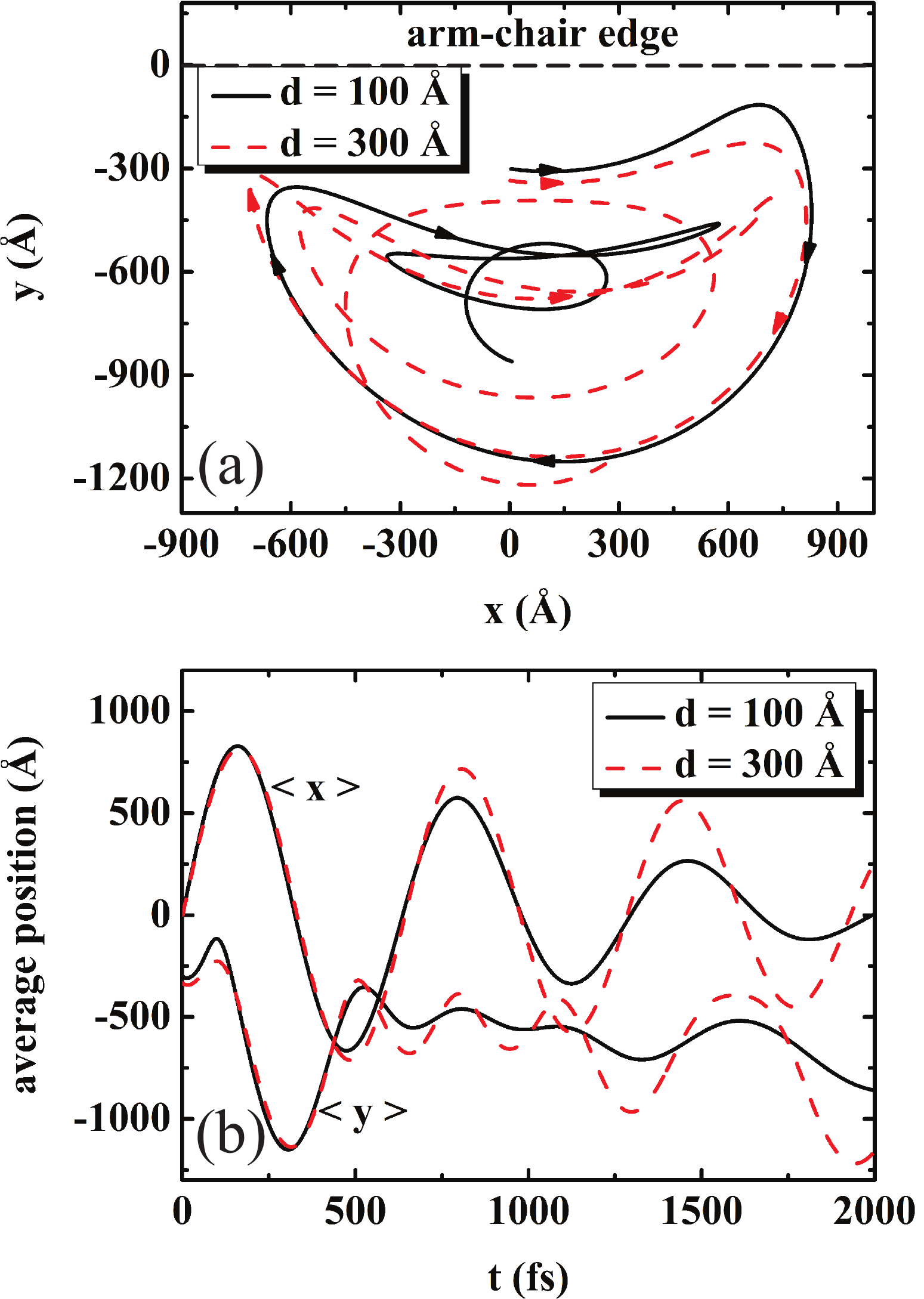}}
\caption{(Color online) (a) Trajectories drawn by $\langle x \rangle$ and $\langle y \rangle$ for a $t = 2000$ fs time evolution of a wavepacket, which propagates close to the upper (armchair) border of a bent rectangular graphene sample, for two values of wavepacket width $d$. The radius of the circular distortion is $R$ = 10$^4$ \AA\,, corresponding to an almost uniform $\approx$ 5 T pseudo-magnetic field. The horizontal dashed line represents the upper edge of the sample. (b) Average values of the wavepacket position $\langle x \rangle$ and $\langle y \rangle$ as a function of time for the trajectories drawn in (a).} \label{fig:trajstrain}
\end{figure}

For a circularly strained graphene flake, like that sketched in Fig. 1(b), electrons in the sample behave as if they were in an almost uniform magnetic field perpendicular to the plane. In order to produce such a strain, our 1801 $\times$ 2000 atoms sample is distorted into a semi-circle of radius $R = 10^4$ \AA\,, leading to a pseudo-magnetic field $\approx 5$ T, \textit{i.e.} close to the value considered for the external magnetic field in Fig. 4. The presence of such a pseudo-magnetic field when electrons move close to the edge are expected to result in skipping orbits, similar to those in Fig. 4. Surprisingly, Fig. 5(a) shows this is not really the case: after performing a semi-circular trajectory due to the Lorentz force coming from the pseudo-magnetic field, the packet, which started in the K valley, is reflected by the upper armchair edge and scatters to the K' valley, where the pseudo-magnetic field points in the opposite direction. The semi-circular trajectory now travels in the opposite direction until the packet reaches the edge again, being scattered back to its former Dirac cone at the K valley. This procedure occurs several times until the packet is so strongly dispersed that it, eventually, does not reach any of the edges, and performs only circular trajectories in the middle of the graphene flake. The series of reflections by the armchair border obtained in the strained case suggests the existence of a quasi-bound state at this edge, which is clearly seen by the time-dependence of the average coordinates $\langle x \rangle$ and $\langle y \rangle$, shown in Fig. 5(b). As time elapses, both $\langle x \rangle$ and $\langle y \rangle$ simply oscillate around $x = 0$ and close to the upper border of the sample, respectively. Notice that differently from Fig. 4(a), the lateral and bottom borders of the panel in Fig. 5(a) do not match the edges of the sample, in order to help the visualization of the trajectory, which in this case is localized in a small region of the sample. One can also observe that the results for different wavepacket widths $d$ are qualitatively the same, differing only by the amplitudes of the $\langle x \rangle$ and $\langle y \rangle$ oscillations in time.

\begin{figure}[!h]
\centerline{\includegraphics[width=\linewidth]{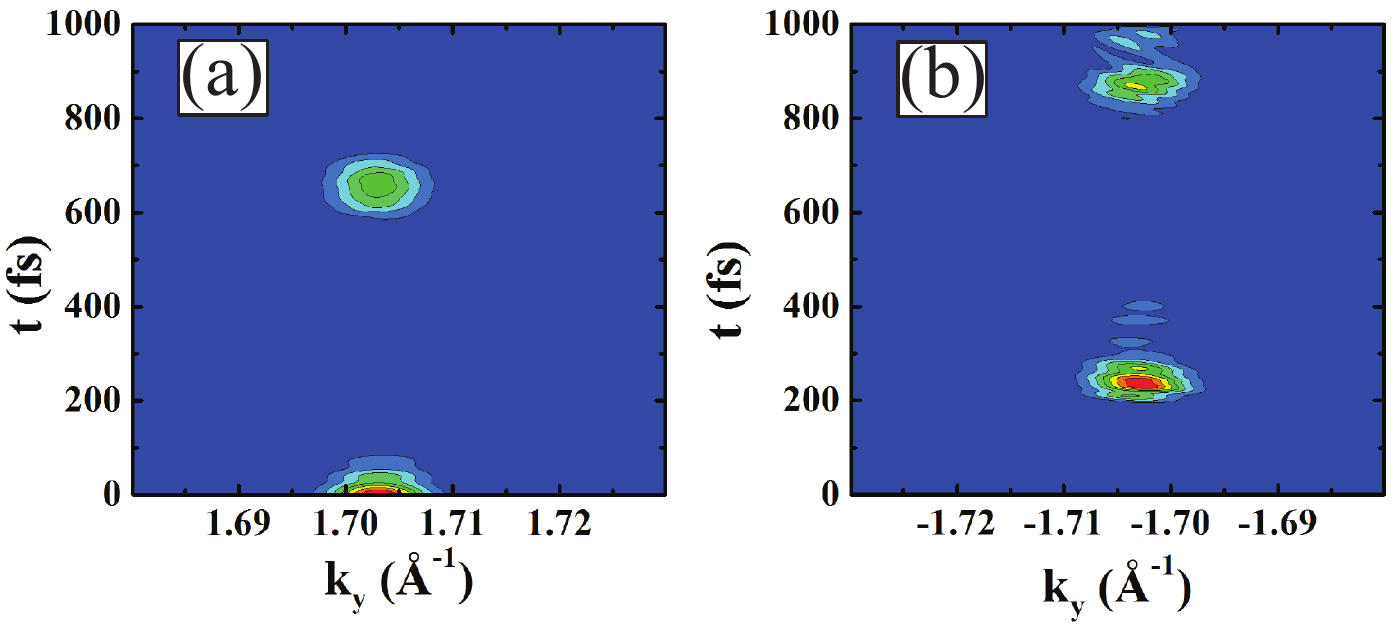}}
\caption{(Color online) Contour plots of the time evolution of the wavepacket in reciprocal space corresponding to the propagation shown in Fig. 5, \textit{i.e.} for a strained graphene sample. The Fourier transform of the wavepacket is taken in the vicinity of the (a) K and (b) K' points of the reciprocal space illustrated in Fig. 3(a), along the $k_x = -0.06$\AA\,$^{-1}$ ($k_x = 0.045$\AA\,$^{-1}$) vertical axis for K (K').} \label{fig:scatteringstrain}
\end{figure}

The sequence of scatterings between K and K' valleys suggested by the trajectories observed in Fig. 5 is confirmed by a direct observation of the wavefunction in reciprocal space. This is illustrated by Fig. 6, which shows the Fourier-transformed wavefunction (contour plots) as a function of the vertical component of the wave vector $k_y$ as function of time $t$, similar to Figs. 3 (c) and (d). Since the initial wavepacket in this case is at $\vec k = (-0.06 $\AA\,$^{-1}, 4\pi\big/3\sqrt{3}a)$, in the vicinity of the K point, the $k_y$ in Fig. 6(a) is taken for a fixed $k_x = -0.06$ \AA\,$^{-1}$, i.e. in the center of the initial wavepacket. One clearly sees that the peak of the wavepacket in reciprocal space oscillates between the K (a) and K' (b) regions in Fig. 6 as time elapses, as a consequence of the successive inter-valley scatterings by the upper armchair edge of the system, as observed in Fig. 5. Indeed, the first peak in the K' cone (Fig. 6(b)), for instance, starts to appear at $t \approx 200$ fs, which is the same time when $\langle x \rangle$ and $\langle y \rangle$ start to decrease in Fig. 5(b), after the wavepacket is scattered by the edge for the first time. It is important to point out that we had to take the $k_x = 0.045$ \AA\,$^{-1}$ vertical axis in order to observe the K' propagation in Fig. 6(b), instead of the $k_x = 0.06$ \AA\,$^{-1}$ that would be expected from the value of the wave vector in our initial wavepacket. In fact, one cannot expect that the K and K' points in the strained case remain vertically aligned in reciprocal space, as illustrated in Fig. 3(a), due to the distortion of the Dirac cones caused by the strain. \cite{CastroNeto300T} Also, we observe that the scattered wavepackets are no longer gaussian, as they start to exhibit interference patterns due to the scattering by the edges. Even so, the conclusions drawn from the results in Fig. 6 are not affected by this fact, while Fig. 6 gives us a clear demonstration that the successive K to K' scatterings are indeed strongly related to the non-propagating edge states found for the strained case in Fig. 5.

\section{Probability density current calculations}

The trajectories illustrated in Figs. 4 and 5 have a direct effect on the probability density currents, which are numerically calculated based on the method developed in Ref. \onlinecite{Current}. Since we can define the probability current $j$ in terms of the continuity equation, then the discrete current centered on site $n$ can be written as
\begin{equation}\label{eq.initialcurrent}
j_n - j_{n+1} = a\frac{\partial}{\partial t}\rho_{n,n},
\end{equation}  
where $\rho_{n,n}=\langle n\vert\hat{\rho}\vert n\rangle$ are the matrix elements of the density matrix operator $\hat \rho = \vert \Psi \rangle \langle \Psi \vert $, and the time derivative is determined by the equation of motion for $\hat{\rho}$ 
\begin{equation}
\frac{\partial}{\partial t}\rho_{nn}=\frac{i}{\hbar}\sum_m \left(\Psi_n\Psi^{*}_{m}H_{mn}-H_{nm}\Psi_m\Psi^{*}_{n}\right)
\end{equation}
where $\Psi_n = \langle n\vert\Psi\rangle$. We will limit ourselves to the case of nearest-neighbor interaction, \textit{i.e.} $H_{n,m}=0$ when $\vert m - n\vert > 1$, from which we obtain
\begin{eqnarray}
\frac{\partial}{\partial t}\rho_{nn}=\frac{i}{\hbar} \left[\left(\Psi_n\Psi^{*}_{n+1}H_{n+1,n}-H_{n,n+1}\Psi_{n+1}\Psi^{*}_{n}\right) \right] \nonumber \\
+ \frac{i}{\hbar} \left[\Psi_n\Psi^{*}_{n-1}H_{n-1,n}-H_{n,n-1}\Psi_{n-1}\Psi^{*}_{n}\right],
\end{eqnarray}
which is easily rewritten in the form
\begin{eqnarray}\label{eq.finalcurrent}
\frac{\partial}{\partial t}\rho_{nn}=-\frac{2}{\hbar} \Im \left[\Psi_n\Psi^{*}_{n+1}H_{n+1,n}\right] \nonumber \\
+ \frac{2}{\hbar} \Im \left[\Psi_{n-1}\Psi^{*}_{n}H_{n,n-1}\right].
\end{eqnarray}
By comparing Eqs. (\ref{eq.initialcurrent}) and (\ref{eq.finalcurrent}), one easily identifies the local current in $n$ as
\begin{eqnarray}\label{eq.finalcurrent2}
j_{n}=\frac{2a}{\hbar}\Im \left[\Psi^{*}_{n}\Psi_{n-1}H_{n,n-1}\right].
\end{eqnarray}

Notice that Eq. (\ref{eq.finalcurrent2}) was developed without taking into account any specific lattice and the presence of magnetic fields. However, a generalization to arbitrary discrete lattice is straightforward, and the presence of a magnetic field is included simply by the Peierls substitution of the hopping parameters. \cite{Governale} As graphene is a hexagonal lattice, the current components in $x$ and $y$ directions have different forms and are site dependent. Defining the sites location through their line ($n$) and column ($m$) positions in the lattice (see Ref. \onlinecite{AndreyStrain}), one obtains
\begin{eqnarray}
j_x(n,m) = \pm \frac{a}{\hbar} \left\{2\Im\left[\Psi_{n,m}\Psi^{*}_{n,m\pm 1}\tau_{n,m\pm 1}\right] \right.  \nonumber\\
 \left. -\Im\left[\Psi_{n,m}\Psi^{*}_{n-1,m}\tau_{n-1,m}\right] -\Im\left[\Psi_{n,m}\Psi^{*}_{n+1,m}\tau_{n+1,m}\right] \right\}
\end{eqnarray}  
and
\begin{eqnarray}
j_y(n,m) = \frac{\sqrt{3}a}{\hbar}\left\{\Im\left[\Psi_{n,m}\Psi^{*}_{n+1,m}\tau_{n+1,m}\right]\right. \nonumber\\
\left. - \Im\left[\Psi_{n,m}\Psi^{*}_{n-1,m}\tau_{n-1,m}\right]\right\}
\end{eqnarray}  
where the $\mp$ sign in $j_x$ will be positive (negative) if the ($n,m$)-site belongs to the sublattice A (B), and $\tau_{n,m}$ is the hopping parameter which, in the presence of a magnetic field, includes an additional phase according to the Peierls substitution $\tau_{n,m}\rightarrow\tau_{n,m}\exp{\left[i\frac{e}{\hbar}\int^{n}_m \vec{A}\cdot d\vec{l}\right]}$, where $\vec{A}$ is the vector potential describing the magnetic field. 

\begin{figure}[!h]
\centerline{\includegraphics[width=\linewidth]{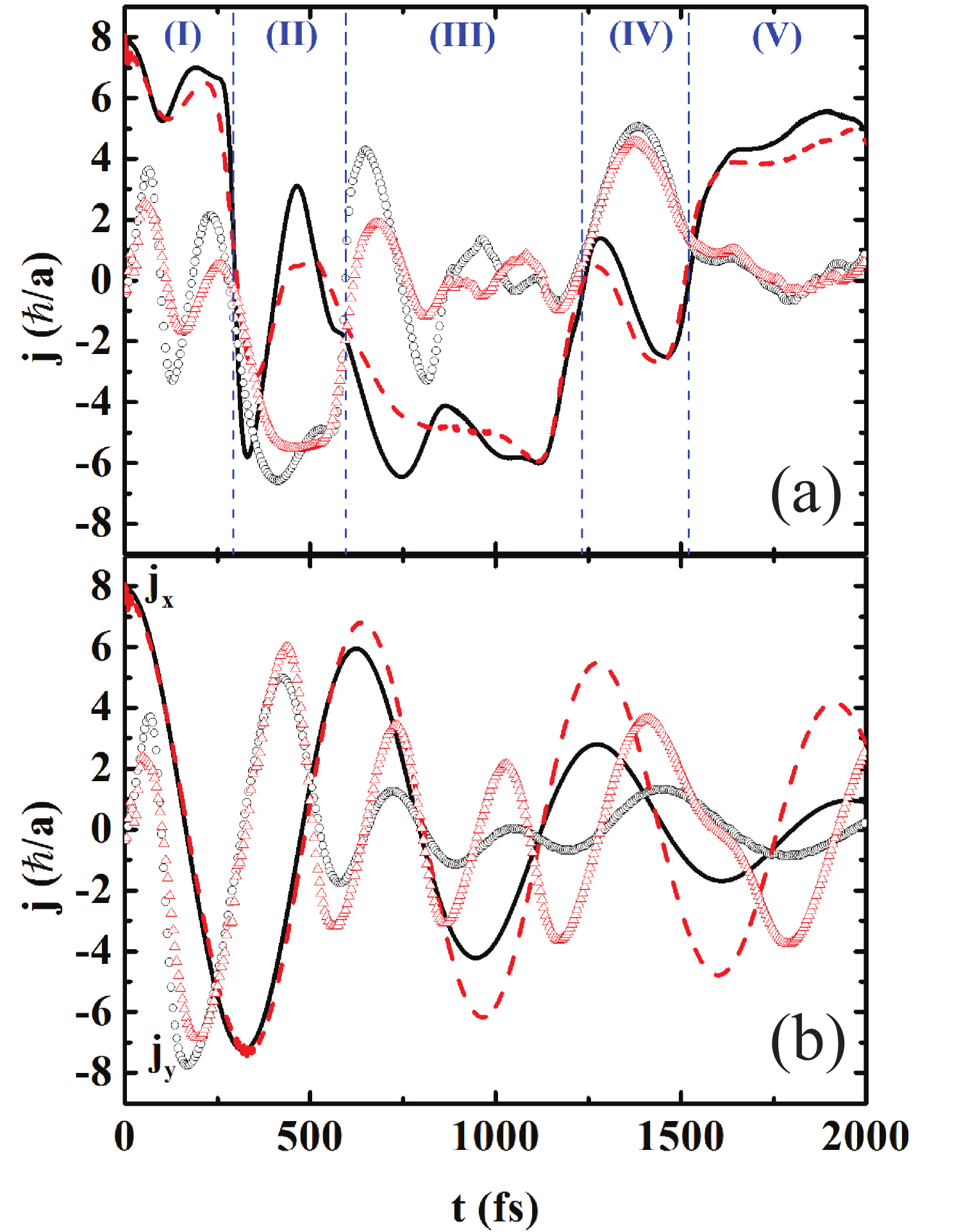}}
\caption{(Color online) Integrated probability density currents as a function of time for the situations proposed in Figs. 4 and 5, namely, (a) for an unstrained graphene sample in the presence of an external 5 T magnetic field, and (b) in a circularly bent graphene sample, which produces an almost uniform $\approx 5$ T pseudo-magnetic field. The curves (symbols) represent the component of the current in the $x$ ($y$)-direction, \textit{i.e.} $j_x$ ($j_y$). Two different values of wavepacket width are considered: $d = 100$ \AA\, (black solid - circles) and 300 \AA\, (red dashed - triangles) The regions delimited in (a) are the same as in Fig. 4.} \label{fig:current}
\end{figure}

The probability density currents calculated by Eq. (\ref{eq.finalcurrent2}) are integrated in space and plotted as a function of time in Fig. 7 for the situations proposed in Figs. 4 and 5, \textit{i.e.} (a) in the presence of an external 5 T magnetic field, and (b) in a strained graphene sample, which produces an almost uniform $\approx 5$ T pseudo-magnetic field. As in the other results discussed previously, the results obtained for the two different values of wavepacket width considered in this case, $d = 100$ \AA\, (black solid - circles) and $d = 300$ \AA\, (red dashed - triangles), exhibit similar qualitative features, differing only in a quantitative way. In Fig. 7(a), for an external field, one observes a total current flow in the $x$-direction oscillating around a positive value in the region I, whereas $j_y$ oscillates around zero in this region. This is a manifestation of the propagation of the wavepacket through the upper edge of the sample, by means of skipping orbits, as illustrated in Fig. 4(a). What follows can also be understood by analyzing Fig. 4(a): In regions II and IV (III and V), where the wavepacket propagates along the vertical (horizontal) edges, the component of the current in the $y$($x$)-direction oscillates around a non-zero value, indicating an electron propagation through the sample by the skipping orbits mechanism. This is not the case when we consider a strain-induced pseudo-magnetic field: Fig. 7(b) shows that both $j_x$ and $j_y$ always oscillating around zero, confirming that there is no net current in the system and that the skipping orbits near the armchair edge in this case are non-propagating states.

The results found in our work have observable consequences in experiments. For example, the edge propagation of electrons through skipping orbits in an ordinary system under an external magnetic field plays an important role in electron transport in the direction parallel to the edge. \cite{skipping2} Our results demonstrate that these skipping orbits are still present in a graphene ribbon under an external magnetic field, but they are not observed in the case of pseudo-magnetic fields in the direction parallel to an armchair border of graphene. This is a clear example that the pseudo-magnetic field has different consequences as compared to a real magnetic field. Therefore, in a strained armchair graphene ribbon, edge electrons should not propagate along the ribbon, so that the transport in these systems must be dominated only by electrons propagating far from the edge. Moreover, the non-propagating state found at the armchair edge of a strained sample is a consequence of periodic inter-valley scattering processes, and this type of scattering has an important effect on Raman spectroscopy. \cite{Raman} Therefore, the successive electron reflections at the armchair edge of a strained sample would manifests itself as an intense peak in Raman experiments taken close to the border of the graphene sample.

\section{Conclusions}

In summary, we investigated the reflection of a wavepacket on zigzag and armchair edges of a graphene ribbon. Our results demonstrate the scattering of the wavepacket from K to K' Dirac cones in the case of armchair edges, whereas scattering from positive to negative average momentum inside the same cone is observed in the zigzag case, which is in agreement with predictions from mean field (Dirac) theory of graphene and with recent experimental results. \cite{Park} In the presence of an external magnetic field, skipping orbits are observed. However, for a strain induced pseudo-magnetic field, our numerical results demonstrate that the incoming and scattered wavepackets perform orbits in opposite directions in the armchair case. This effect is easy to be understood if one considers the combination between two events, both already predicted by the Dirac theory of graphene: (i) the K to K' scattering by armchair edges and (ii) the opposite sign of the pseudo-magnetic field in the different cones. This result points directly to the possibility of observing non-propagating edge states in an armchair terminated strained graphene sample under pseudo-magnetic fields, which is completely different from the external magnetic fields case, where the skipping orbit states are always propagating. The effects predicted by our theoretical work are expected to have important consequences in future experiments on strained graphene samples. 

\acknowledgments
Discussions with E. B. Barros are gratefully aknowledged. This work was supported by the Brazilian Council for Research (CNPq), the Flemish Science Foundation (FWO-Vl), the ESF-EuroGRAPHENE (project CONGRAN) and the Bilateral program between Flanders and Brazil.


\begin{thebibliography}{apsrev}

\bibitem{CastroNetoReview} A. H. Castro Neto, F. Guinea, N. M. R. Peres, K. S. Novoselov and A. K. Geim, Rev. Mod. Phys. \textbf{81}, 109 (2009).

\bibitem{Katsnelson} M. I. Katsnelson, K. S. Novoselov, and A. K. Geim, Nature Phys. \textbf{2}, 620 (2006).

\bibitem{Maksimova} G. M. Maksimova, V. Ya. Demikhovskii, and E. V. Frolova, Phys. Rev. B \textbf{78}, 235321 (2008).

\bibitem{Yang} H. Yang, A. J. Mayne, M. Boucherit, G. Comtet, G. Dujardin, and Young Kuk, Nano Lett. 10, 943 (2010).

\bibitem{Gunlycke} D. Gunlycke and C. T. White, Phys. Rev. Lett. \textbf{106}, 136806 (2011)

\bibitem{Massoud} M. Ramezani Masir, A. Matulis, and F. M. Peeters, Phys. Rev. B \textbf{84}, 245413 (2011).

\bibitem{Chen} Jian-Hao Chen, W. G. Cullen, C. Jang, M. S. Fuhrer, and E. D. Williams, Phys. Rev. Lett. \textbf{102}, 236805 (2009).

\bibitem{TheoryReconstruction} J. A. M. van Ostaay, A. R. Akhmerov, C. W. J. Beenakker, and M. Wimmer, Phys. Rev. B \textbf{84}, 195434 (2011).

\bibitem{ExpReconstruction1} P. Koskinen, S. Malola, and H. H\"akkinen, Phys. Rev. Lett. \textbf{101}, 115502 (2008).

\bibitem{ExpReconstruction2} P. Koskinen, S. Malola, and H. H\"akkinen, Phys. Rev. B \textbf{80}, 073401 (2009).

\bibitem{ExpReconstruction3} P. Rakyta, A. Korm\'anyos, J. Cserti and P. Koskinen, Phys. Rev. B \textbf{81}, 115411 (2010).

\bibitem{Akhmerov} A. R. Akhmerov and C. W. J. Beenakker, Phys. Rev. B \textbf{77}, 085423 (2008).

\bibitem{Mohammad} M. Zarenia, A. Chaves, G. A. Farias, and F. M. Peeters, Phys. Rev. B \textbf{84}, 245403 (2011). 

\bibitem{Park} C. Park, H. Yang, A. J. Mayne, G. Dujardin, S. Seo, Y. Kuk, J. Ihm, and G. Kim, PNAS \textbf{108}, 18622 (2011).

\bibitem{CastroNetoStrain} V. M. Pereira and A. H. Castro Neto, Phys. Rev. Lett. \textbf{103}, 046801 (2009).

\bibitem{new1} E. J. G. Santos, A. Ayuela, S. B. Fagan, J. Mendes Filho, D. L.
Azevedo, A. G. Souza Filho, and D. S\'anchez-Portal, Phys. Rev. B \textbf{78}, 195420 (2008).

\bibitem{new2} V. M. Pereira and A. H. Castro Neto, Phys. Rev. Lett. \textbf{103}, 046801 (2009).

\bibitem{new3} A. T. N'Diaye, R. van Gastel, A. J. Martinez-Galera, J.
Coraux, H. Hattab, D. Wall, F. -J. M. Zu Heringdorf, M. H. Hoegen, J. M.
Gomez-Rodr\'iguez, B. Poelsema, C. Busse, and T. Michely, New J. of
Phys. \textbf{11}, 113056 (2009).

\bibitem{new4} F. M. D. Pellegrino, G. G. N. Angilella, and R. Pucci,
Phys. Rev. B \textbf{81}, 035411 (2010).

\bibitem{new5} C. Metzger, S. Reemi, M. Liu, S. V. Kusminskiy, A. H.
Castro Neto, A. K. Swan, and B. B. Goldberg, Nano Lett. \textbf{10}, 6
(2010).

\bibitem{new6} G. Cocco, E. Cadelano, and L. Colombo, Phys. Rev. B
\textbf{81}, 241412 (2010).

\bibitem{new7} F. M. D. Pellegrino, G. G. N. Angilella, and R. Pucci, Phys. Rev. B
\textbf{82}, 115434 (2010).

\bibitem{new8} O. Cretu, A. V. Krasheninnikov, J. A. Rodr\'iguez-Manzo, L.
Sun, R. M. Nieminen, and F. Banhart, Phys. Rev. Lett. \textbf{105}, 196102 (2010).

\bibitem{new9} L. Covaci and F. M. Peeters, Phys. Rev. B \textbf{84}, 241401 (2011).

\bibitem{new10} E.J.G. Santos, A. Ayuela, and D. Sanchez-Portal, J.
Phys. Chem. C \textbf{116}, 1174 (2012).

\bibitem{Guinea} F. Guinea, M. I. Katsnelson, and A. K. Geim, Nature Physics \textbf{6}, 30 (2010).

\bibitem{CastroNeto300T} N. Levy, S. A. Burke, K. L. Meaker, M. Panlasigui, A. Zettl, F. Guinea, A. H. Castro Neto, and M. F. Crommie, Science \textbf{30}, 544 (2010).

\bibitem{GuineaStrain} F. Guinea, A. K. Geim, M. I. Katsnelson, and K. S. Novoselov, Phys. Rev. B \textbf{81}, 035408 (2010). 

\bibitem{AndreyStrain} A. Chaves, L. Covaci, Kh. Yu. Rakhimov, G. A. Farias, and F. M. Peeters, Phys. Rev. B \textbf{82}, 205430 (2010).

\bibitem{Rusin} T. M. Rusin and W. Zawadzki, Phys. Rev. B \textbf{78}, 125419 (2008).

\bibitem{skipping} R. J. Haug, Semicond. Sci. Technol. \textbf{8}, 131 (1993).

\bibitem{skipping2} T. Usuki, M. Saito, M. Takatsu, R. A. Kiehl, and N. Yokoyama, Phys. Rev. B \textbf{52}, 8244 (1995).

\bibitem{Current} E. A. de Andrada e Silva, Am. J. Phys. \textbf{60}, 8 (1992).

\bibitem{Governale} M. Governale and C. Ungarelli, Phys. Rev. B
\textbf{58}, 7816 (1998).

\bibitem{Raman} L. M. Malard, M. A. Pimenta, G. Dresselhaus, and M. S. Dresselhaus, Physics Reports \textbf{473}, 51 (2009).

\end{thebibliography}
\end{document}